# Magnetic Excitations in Strained Infinite-layer Nickelate PrNiO$_2$

Qiang Gao[1,#], Shiyu Fan[2,#], Qisi Wang[3,#], Jiarui Li[4], Xiaolin Ren[1,5], Izabela Biało[3,6], Annabella Drewanowski[3], Pascal Rothenbühler[3], Jaewon Choi[7], Yao Wang[8], Tao Xiang[1,5,9], Jiangping Hu[1,5], Ke-Jin Zhou[7], Valentina Bisogni[2], Riccardo Comin[4], J. Chang[3], Jonathan Pelliciari[2,*], X. J. Zhou[1,5,9,10,*], and Zhihai Zhu[1,5,10,*]

[1]Beijing National Laboratory for Condensed Matter Physics,
Institute of Physics, Chinese Academy of Sciences, Beijing 100190, China

[2]National Synchrotron Light Source II,
Brookhaven National Laboratory, Upton, New York 11973, USA

[3]Physik-Institut, Universität Zürich,
Winterthurerstrasse 190, CH-8057 Zürich, Switzerland

[4]Department of Physics, Massachusetts Institute of Technology,
Cambridge, Massachusetts 02139, USA

[5]University of Chinese Academy of Sciences,
Beijing 100049, China

[6]AGH University of Science and Technology,
Faculty of Physics and Applied Computer Science, 30-059 Kraków, Poland

[7]Diamond Light Source, Harwell Campus,
Didcot OX11 0DE, United Kingdom

[8]Department of Physics and Astronomy,
Clemson University, Clemson, SC 29631, USA

[9]Beijing Academy of Quantum Information Sciences, Beijing 100193, China

[10]Songshan Lake Materials Laboratory, Dongguan 523808, China

[#]These authors contributed equally to the present work.

(Dated: Aug 11, 2022)

[*]To whom correspondence should be addressed.

Emails: pelliciari@bnl.gov, XJZhou@iphy.ac.cn, zzh@iphy.ac.cn

**Strongly correlated materials often respond sensitively to the external perturbations. In the recently discovered superconducting infinite-layer nickelates, the superconducting transition temperature can be dramatically enhanced via only ~1% compressive strain-tuning enabled by substrate design. However, the root of such enhancement remains elusive. While the superconducting pairing mechanism is still not settled, magnetic Cooper pairing - similar to the cuprates has been proposed. Using resonant inelastic x-ray scattering, we investigate the magnetic excitations in infinite-layer $PrNiO_2$ thin films for different strain conditions. The magnon bandwidth of $PrNiO_2$ shows only marginal response to strain-tuning, in sharp contrast to the striking enhancement of the superconducting transition temperature $T_c$ in the doped superconducting samples. These results suggest the enhancement of $T_c$ is not mediated by spin excitations and thus provide important empirics for the understanding of superconductivity in infinite-layer nickelates.**

High-temperature superconductivity continues to be a challenging topic in studies of correlated quantum matter since multiple electronic phases emerge in proximity to each other, masking the leading interaction for the electron pairing. The newly discovered superconducting infinite-layer nickelates have provided a new platform to study unconventional superconductivity [1]. A question that has soon arisen as a central one for these systems is to what extent they are analogues of cuprate superconductors. Understanding the similarities and distinctions between these two families of materials may help bring to light new aspects of high-$T_c$ superconductivity, and in particular the pairing mechanism [2]. Recent experiments have revealed remarkable differences between these two classes of materials. For instance, the parent compounds of infinite-layer nickelates are Mott-Hubbard insulators, while cuprates are charge-transfer insulators according to the Zaanen-Sawatzky-Allen (ZSA) scheme [3–6]; unlike the cuprates, the rare-earth spacer layers in infinite-layer nickelates hybridize with Ni $3d$ orbitals, leading to 5$d$ metallic states at the Fermi level [3, 4]. Despite these differences, infinite-layer nickelates share several general characteristics with cuprates, including a linear temperature dependence of resistivity for the normal state [7], a dome-like shape for $T_c$ as a function of doping in the phase diagram [8–13], a sizable magnetic exchange interaction [14-17], the charge density wave instabilities [16-18], and a possible $d$-wave superconducting gap [19-21]. These properties corroborate that the superconductivity in infinite-layer nickelates is unconventional.

An approach to tackling the pairing mechanism is to directly manipulate $T_c$ with controllable knobs, and simultaneously examine the response of the bosonic excitations to these perturbations. By examining the coupling between low-energy excitations and external tuning parameters, one may identify the leading interaction channels accounting for the superconductivity. The use of different substrates has been proposed as a pathway for tuning $T_c$ in thin films of infinite-layer nickelates [1]. However, it is highly challenging to obtain superconducting films on substrates other than SrTiO$_3$ (STO). Recent experiments have shown a dramatic boost of $T_c$ with more than 40% for the superconducting infinite-layer nickelates via only ~1% compressive strain, achieved by growing the films on (LaAlO$_3$)$_{0.3}$(Sr$_2$TaAlO$_6$)$_{0.7}$ (LSAT) instead of STO [7, 22]. This strain-induced enhancement of $T_c$ has also been observed in La$_{2-x}$Sr$_x$CuO$_4$ thin films grown on different substrates [23–25], where the increase of $T_c$ is attributed to strengthening the magnetic exchange interaction by compressive strain [26]. Although long-range magnetic ordering has not been found in infinite-layer nickelates to date, resonant inelastic x-ray scattering (RIXS) studies on Nd$_{1-x}$Sr$_x$NiO$_2$ have revealed propagating spin excitations, resembling the ones of the spin-1/2 antiferromagnet (AFM) on the square lattice, with a large spin exchange energy ~60 meV in the parent compound [14]. Presently, how the magnetic excitations couple to strain, thus to the lattice, is still unexamined. In infinite-layer nickelates, difficulties arise from the limited scattering volume precluding inelastic neutron scattering investigations.

Here, using high-resolution Ni $L_3$-edge RIXS, we unveiled the role of strain on magnetic excitations in infinite-layer $PrNiO_2$ thin films grown on STO and LSAT, respectively. The epitaxial strain imposed by the substrates shows marginal influence on the bandwidth of magnon dispersion, which is in remarkable contrast to the dramatic enhancement of $T_c$ under similar strain-tuning for superconducting films. These findings suggest that the energy scale of spin fluctuations is not correlated with $T_c$, at odds with what has been reported in cuprates [26-28]. These results hint that the enhancement of $T_c$ is not mediated by magnetic interactions and thus provide important implications for the determination of the electron pairing mechanism in superconducting infinite-layer nickelates.

Figure 1a shows a schematic of the scattering geometry of our RIXS measurements. It has been shown that the magnetic excitations in cuprates can be clearly detected with RIXS by using either grazing-in geometry with σ (linear vertical) incident light polarization or grazing-out geometry with π (linear horizontal) polarized incident photons [29]. We adopted the former for the majority of the data in the present study. The x-ray absorption spectra (XAS) of the $PrNiO_2$ film grown on STO display a much stronger absorption peak in σ than π polarization at the Ni $L_3$ edge ($2p^63d^9 - 2p^53d^{10}$ transition) (see Fig. 1b). This linear-dichroism reflects the $d_{x^2-y^2}$ symmetry of the $3d$ hole in $PrNiO_2$. In the $PrNiO_2$ film on LSAT, the linear-dichroism is lower as shown in Fig. 1c. A possible extrinsic reason is due to the strong signal at ~850 eV that is associated with the La $M_4$ edge ($3d - 4f$ transition) in the LSAT substrate.

Moving to the RIXS spectra, the high energy ($dd$) excitations provide valuable information regarding the local configurations of the $3d$ orbitals in Ni ions which is determined by the symmetry of crystal field. In infinite-layer nickelates, the expected $D_{4h}$ crystal field leads to the splitting of the Ni $3d$ orbitals with the $d_{x^2-y^2}$ at the highest energy, followed in sequence by the $d_{xy}$, $d_{yz/xz}$, and $d_{3z^2-r^2}$, as illustrated in Fig. 1b. We show in Figs. 1e and 1f the $dd$ excitations in the RIXS spectra of $PrNiO_2$ on STO and LSAT, respectively. Both spectra exhibit four major features marked by the shaded areas. The spectral peaks in the energy-loss range of 1 - 4 eV correspond to the crystal field splitting as illustrated in Fig. 1b, and the peak at ~0.7 eV arises from the hybridization between Ni and Pr ions, which is similar to studies on $ANiO_2$ (A= La, Nd) [3, 14, 16-18, 30]. Overall, the dominant spectral features assigned to the *dd* excitations are similar for both samples. As denoted by the blue dashed lines, the peaks assigned to the transition to $d_{xy}$ orbital are comparable for both samples regarding the peak center of mass positions as well as line shapes; the peak assigned to the transition to $d_{yz/xz}$ orbital moves toward lower energy for $PrNiO_2$/LSAT compared with $PrNiO_2$/STO. The discernible changes of *dd* excitations demonstrate that the electronic structures of $PrNiO_2$ are effectively modified by epitaxial strain.

In Fig. 2, we show the low-energy loss RIXS spectra for $PrNiO_2$/STO and $PrNiO_2$/LSAT along high symmetry directions (*h*, 0) and (*h*, *h*) in momentum space. The spectra of $PrNiO_2$/STO consist of three major features composed of an elastic peak at zero energy loss, a clear excitation at ~60 meV, and a broad peak in the range of 100-400 meV. The peak at ~60 meV represents the phonon excitation, reminiscent of the ~70 meV phonon mode that prevails in cuprates [14, 31]. The broad peak is ascribed to the magnetic excitations as it disperses as a function of momentum, resembling what was observed in $NdNiO_2$ and $LaNiO_2$ [14, 16, 17]. In $PrNiO_2$/LSAT the elastic peak is more prominent probably owing to the contribution from the La $M_4$ edge in the substrate. Nevertheless, the excitations can still be reliably extracted by fitting the RIXS spectra to a combination of a Voigt function for the elastic peak, a Gaussian function for the phonon, a damped harmonic oscillator (DHO) to account for the magnetic excitation, and a smoothly varying background. The DHO function $\chi''(q, \omega)$ is given by

$$\chi''(q,\omega) = \frac{\gamma_q \omega}{\left(\omega^2 - \varepsilon_q^2\right)^2 + 4\gamma_q^2 \omega^2} \qquad (1)$$

where $\varepsilon_q$ is the undamped mode energy and $\gamma_q$ is the damping factor [29]. As shown in Fig. 2, the fitting overall describes well the experimental spectra.

To better represent the magnon dispersion, we include in Figs. 3a and 3b the magnetic spectra map of the magnetic excitations after subtracting the elastic peak, phonon peak, and background from the raw data. As shown in Fig. 3a, a clear magnon dispersion can be visualized along both ($h$, 0) and ($h$, $h$) directions for PrNiO$_2$/STO, while in the case of PrNiO$_2$/LSAT (see Fig. 3b), the magnetic excitations appear to be less dispersive along both directions. To directly characterize the response of magnetic excitations to strain-tuning, we show in Figs. 3c-3f the RIXS spectra at the zone boundaries. Here π incident light polarization and grazing-out scattering geometry were used to enhance the magnon intensity. Again, the DHO function was used to fit the magnon peaks. The grey dashed lines represent the magnon energies from the fits, while the blue dashed lines denote the energies of peak maximum. We find the magnons move toward lower energies when strain is applied.

Figure 4a shows the momentum dependence of the fitted values of the magnetic excitation energy $\varepsilon_q$ and the damping factor $\gamma_q$ as defined in Eq. (1) for both samples. Considering the uncertainty for the fitting, the magnon dispersions are comparable in both cases, with an energy maximum close to (0.5, 0) along ($h$, 0) direction, and close to (0.25, 0.25) along ($h$, $h$) direction; they are similar to the magnon dispersions for the spin-1/2 Heisenberg AFM on the square lattice. The magnon bandwidth in PrNiO$_2$/LSAT appears to be slightly reduced compared to that in PrNiO$_2$/STO. The damping factors of the two samples are comparable and vary little along both directions. To model the energy scale of the spin excitations, we fit the extracted magnetic dispersion resorting to linear spin wave theory [32]. The Hamiltonian is given by

$$H = J_1 \sum_{\langle ij \rangle} S_i \cdot S_j + J_2 \sum_{\langle ii' \rangle} S_i \cdot S_{i'} \qquad (2)$$

where $S_i$ is the spin-1/2 operator on site $i$, and $\langle ij \rangle$ ($\langle ii' \rangle$) denotes the nearest neighbours (next nearest neighbours). The best-fit to the spectra yields $J_1$ ($J_2$) = 66.5 (-7.5) meV for PrNiO$_2$/STO, and $J_1$ ($J_2$) = 64 (-5.5) meV for PrNiO$_2$/LSAT, similar to the findings for NdNiO$_2$ on STO [14]. All in all, the spin exchange coupling for PrNiO$_2$/LSAT is nearly equal to that for PrNiO$_2$/STO, suggesting that the in-plane compressive strain of ~1% has a marginal influence on the superexchange coupling $J$.

A major motivation of this study is to examine whether the recent findings of the strain-induced enhancement of $T_c$ pertain to the magnetic excitations in superconducting infinite-layer nickelates. In the RIXS studies on La$_2$CuO$_4$ thin films, the Coulomb and magnetic exchange interactions are strengthened by the compressive strain imposed by the substrates, which may account for the doubling of $T_c$ in the doped films [26]. Empirically, the superexchange $J$ in an insulator is expected to scale with inter-ion distance $a$ by $J \sim a^{-10}$ [33]. This would approximately lead to a ~10% enhancement of $J$ for 1% compressive strain. First-principles calculations predict that the magnon bandwidth increases by 7.8% for a -1% strain [22]. From our results, the energy scale of the spin excitations, which is determined by the superexchange interaction between the planar nearest neighbour Ni spins, seems to be slightly reduced when a ~1% compressive strain is applied. A possible explanation for this difference with respect to La$_2$CuO$_4$ is that there may exist a structural distortion of the Ni-O plane, which would modify the bonding angle of Ni-O-Ni under compressive strain, and hence the superexchange $J$. As illustrated in Fig. 4b, such structural distortion is commonly seen in the precursor phase $R$NiO$_3$ ($R$ is a rare-earth element such as La, Pr, Nd, and Sm) [34, 35]. After the topotactic transformation, the structural distortion may exist in the infinite-layer phase [36]. Similarly, the structural distortion with modified bonding-angles has been proposed to explain the stripe-like charge ordering in La$_4$Ni$_3$O$_8$ and La$_3$Ni$_2$O$_6$ [37]. The possibility of periodically modified bonding-angles may also help understand the recent observations of charge orderings in infinite-layer nickelates [16-18]. Alternatively, the lattice fluctuation induced by the electron-phonon interaction suppresses the effective superexchange $J$, as recently proposed for parent compounds of cuprates [38]. In this scenario, bond-stretching phonons dynamically drag oxygen away from its equilibrium position, as illustrated in Fig. 4c; accordingly, the superexchange strength would be suppressed as the electron-phonon coupling increases under compressive strain. However, such a picture has been demonstrated only for one-dimensional (1D) systems, and whether it can be generalized to 2D systems remains unclear. In any case, this reduced magnetic exchange coupling due to 1% compressive strain in infinite-layer nickelates differs from that in the $La_2CuO_4$ thin films mentioned above [26], suggesting that the dramatic enhancement of $T_c$ in superconducting infinite-layer nickelates may not be dominated by spin interactions.

Our results have important implications for the origin of pairing interactions responsible for superconductivity in infinite-layer nickelates. For unconventional superconductivity, spin fluctuations are the main candidate as the driving force for condensing electrons into pairs [39]. In this scenario, spin interactions primarily set the energy scale for superconductivity, which occurs in close proximity to an antiferromagnetic phase. In the large $U$ limit, $T_c$ is expected to scale with the superexchange $J$ at the mean-field level for the studies on cuprate superconductors [27, 40]. This has also been used to estimate $T_c$ for the potential $LaNiO_3/LaMO_3$ superconductors, where M = Al, Ga, and Ti [41]. According to our results, the bandwidth of the spin excitations does not directly scale with $T_c$ in infinite-layer nickelates, at odds with what has been recently reported for cuprates [26-28]. Noticeably, the observation of phonon modes at ~60 meV, reminiscent of the phonon modes that prevail in various cuprates [31, 42-49], suggests considerable electron-phonon coupling exists in infinite-layer nickelates, though its role in superconductivity remains to be addressed in the future by direct experimental evidence.

## METHODS

**Sample preparation.** Thin films of the precursor phase $PrNiO_3$ with a thickness of ~7 nm were prepared by using pulsed laser deposition (PLD) on (001)-oriented $SrTiO_3$ and LSAT substrates with a 248-nm KrF excimer laser [22]. The infinite-layer phase $PrNiO_2$ was obtained by soft-chemistry reduction process using $CaH_2$ powder. Substrates were preannealed at 900°C with an oxygen partial pressure of 1 × 10$^{-5}$ Torr. During growth, the substrate was kept at 600°C under an oxygen partial pressure of 150 mTorr. After deposition the films were cooled to room temperature at a rate of 5°C per minute in the same oxygen partial pressure. For $CaH_2$ topotactic reduction, the as-grown nickelate films were sealed with 0.1 g $CaH_2$ powder and annealed at a temperature of ~290°C for ~3h. After reduction, the $PrNiO_2$ films were loaded back to the PLD chamber and capped with amorphous $SrTiO_3$ layer of 10 nm at room temperature to protect the films.

**XAS and RIXS measurements.** The X-ray absorption spectroscopy (XAS) measurements at the Ni *L*-edge were performed at the SIX 2-ID beamline at the National Synchrotron Light Source II (NSLS-II), Brookhaven National Laboratory (USA). The spectra were collected by total fluorescence yield (TFY) at 40 K with linear vertical (σ) and horizontal (π) light polarizations. The spectra are normalized to the incident photon flux.

High-resolution Resonant Inelastic X-ray Scattering (RIXS) measurements were mainly performed at the SIX 2-ID beamline of NSLS-II using σ incident light polarization. The energy resolution was set to ΔE = 34 meV (full-width-at-half-maximum) at the Ni $L_3$ edge [50]. Additional RIXS spectra (Fig. 3c-3f) were taken with π incident light polarization at the I21 beamline at the Diamond Light Source, where the energy resolution was set to 39 and 54 meV for measurements on $PrNiO_2$/STO and $PrNiO_2$/LSAT, respectively. The SIX (I21) spectrometer was positioned at the largest scattering angle of 150˚ (154˚) to maximize the in-plane momentum transfer, and the sample temperature was set to 40 (16) K. All RIXS spectra are normalized to the area of the *dd* excitations (400 - 4000 meV).

**ACKNOWLEDGMENTS**

We thank Fu-Chun Zhang, Shiliang Li, Yi Zhou, and Yuan Wan for fruitful discussions. This work was supported in part by the National Natural Science Foundation of China (Grant No. 12074411) and (Grant No. 11888101), the National Key Research and Development Program of China (Grant No. 2016YFA0300300 and 2017YFA0302900), the Strategic Priority Research Program (B) of the Chinese Academy of Sciences (Grant No. XDB25000000) and the Research Program of Beijing Academy of Quantum Information Sciences (Grant No. Y18G06). Work at Brookhaven National Laboratory was supported by the U.S. Department of Energy (DOE) Office of Science under Contract No. DE-SC0012704, Early Career Research Program, and the Laboratory Directed Research and Development project of Brookhaven National Laboratory under Contract No. 21-037. J.L. and R.C. acknowledge support by the Air Force Office of Scientific Research Young Investigator Program under grant FA9550-19-1-0063. Q.W. and J.C. acknowledge support by the Swiss National Science Foundation.

**AUTHOR CONTRIBUTIONS**

Q.G., X.J.Z. and Z.H.Z. conceived the research. Q.G. and X.L.R. prepared and characterized the film samples. J.P., S.Y.F., Q.G., J.R.L., Z.H.Z., R.C., Q.S.W., I.B., A.D., P.R., J. Choi, and J. Chang performed the RIXS experiments with the help of V.B., and K.J.Z.. Q.G., Z.H.Z., and Q.S.W. analyzed the data. Y.W., T.X., and J.P.H. provided theoretical understanding. Q.G. and Z.H.Z. wrote the manuscript with input from all authors.

**DECLARE OF INTEREST**

The authors declare no competing interests.

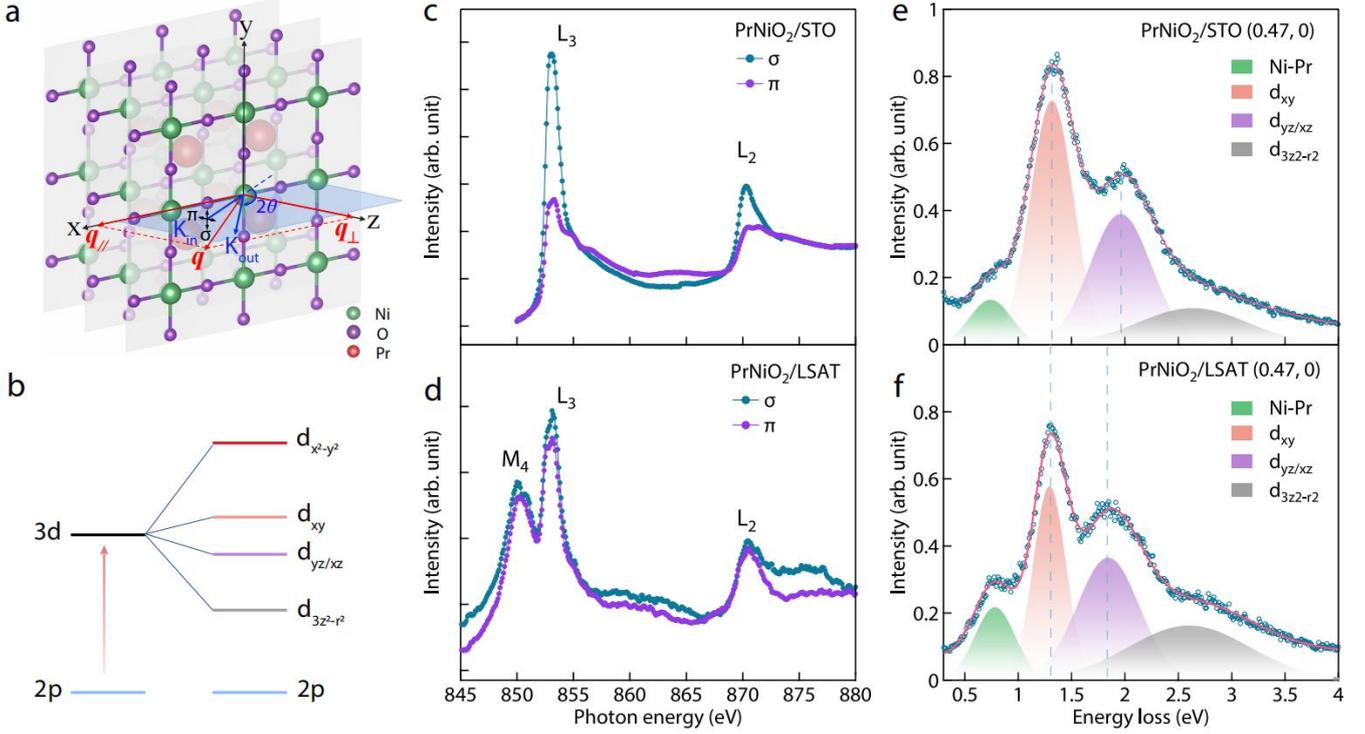

**FIG. 1: X-ray absorption spectra (XAS) and high energy (*dd*) excitations of PrNiO$_2$ films grown on STO and LSAT.** a, Crystal structure of PrNiO$_2$ and scattering geometry of the RIXS experiments. The polarization of the incoming photon is fixed to σ or π, where σ and π represent, respectively, the polarization components, perpendicular and parallel to the scattering plane. The 2θ scattering angle is fixed at 150˚ (or 154˚) to maximize the in-plane momentum transfer, which is tuned by rocking the sample. $q_{//}$ ($q_{\perp}$) refers to the momentum transfer that is parallel (perpendicular) to the nickel-oxide plane, respectively. b, The *d*-level splitting of Ni ion in $D_{4h}$ crystal field. c-d, The XAS of the PrNiO$_2$ films grown on STO and LSAT substrate measured by σ and π polarization. All the XAS measurements were performed with the grazing-in geometry at an incident angle of 10 degrees. e-f, The high energy *dd* excitations of the PrNiO$_2$ films grown on STO and LSAT at a representative momentum, respectively. The blue dashed lines represent the peak positions of the $d_{xy}$ and $d_{yz/xz}$ orbital excitations.

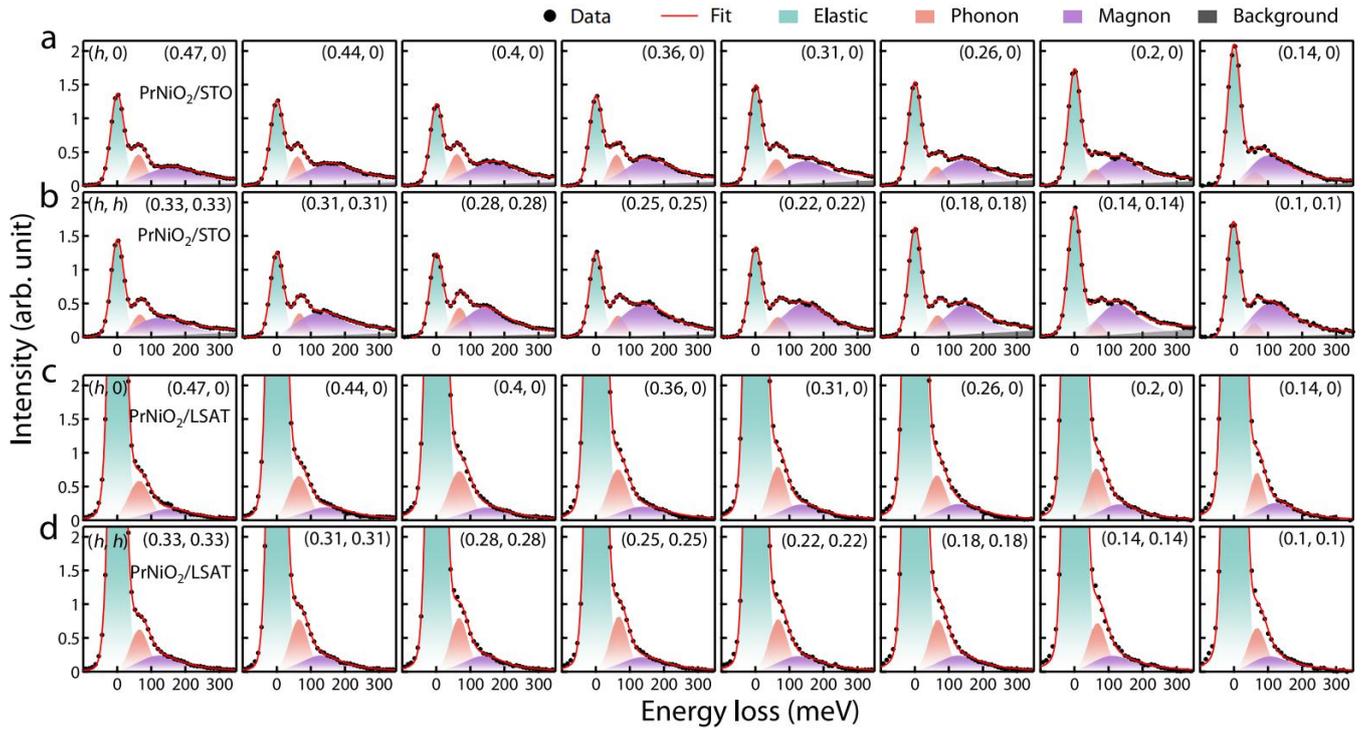

**FIG. 2: Momentum resolved RIXS spectra along high symmetry directions.** a,b, RIXS spectra of the PrNiO$_2$ film grown on STO along ($h$, 0) and ($h$, $h$) directions. c,d, RIXS spectra of the PrNiO$_2$ film grown on LSAT along ($h$, 0) and ($h$, $h$) directions. The filled black circles represent the data and the solid red curves fit the data set, using a combination of an elastic scattering contribution (green), a Gaussian profile for the phonon peak (orange), a DHO function for the magnetic excitation (purple), and background (grey). All the measurements were taken at 40 K.

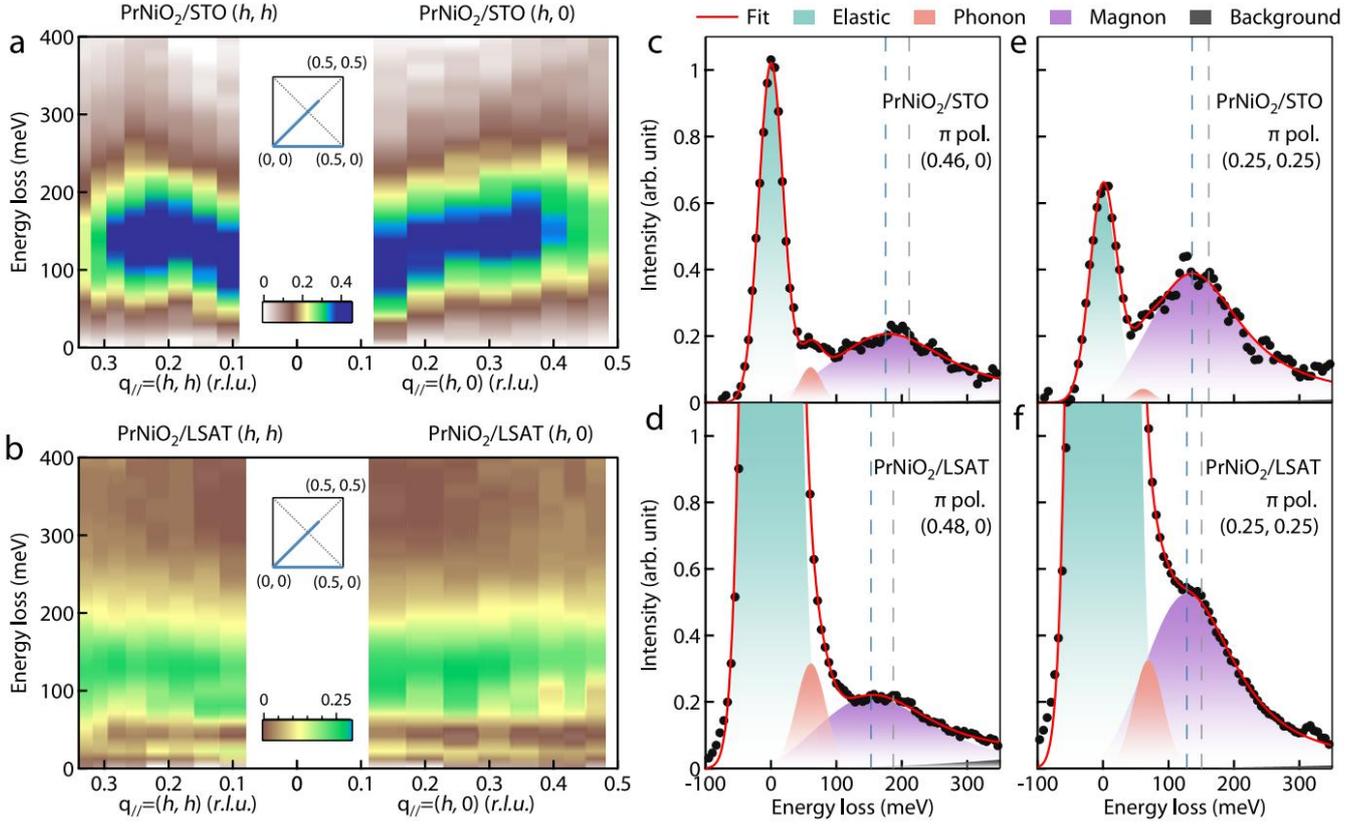

**FIG. 3: Dispersion of the magnetic excitations in PrNiO₂.** a,b, RIXS intensity map of the PrNiO$_2$ films grown on SrTiO$_3$ (a) and LSAT (b) along ($h, h$) and ($h, 0$) directions at 40 K; which were obtained by subtracting the elastic peak, phonon, and the background components for a better visualization. The insets in (a) and (b) show the trajectory in momentum space of the RIXS measurements. c-f, RIXS spectra of the PrNiO$_2$ film grown on STO and LSAT at the zone boundaries measured by π polarization. The filled black circles represent the data and the solid red curves fit the data set, using a combination of an elastic scattering contribution (green), a Gaussian profile for the phonon peak (orange), a DHO function for the magnetic excitation (purple), and a smoothly varying background (grey). The grey dashed lines represent the mode energy ($\varepsilon_q$) of the magnetic excitations obtained by fitting with the DHO function and the blue dashed lines represent the peak energy of the magnon.

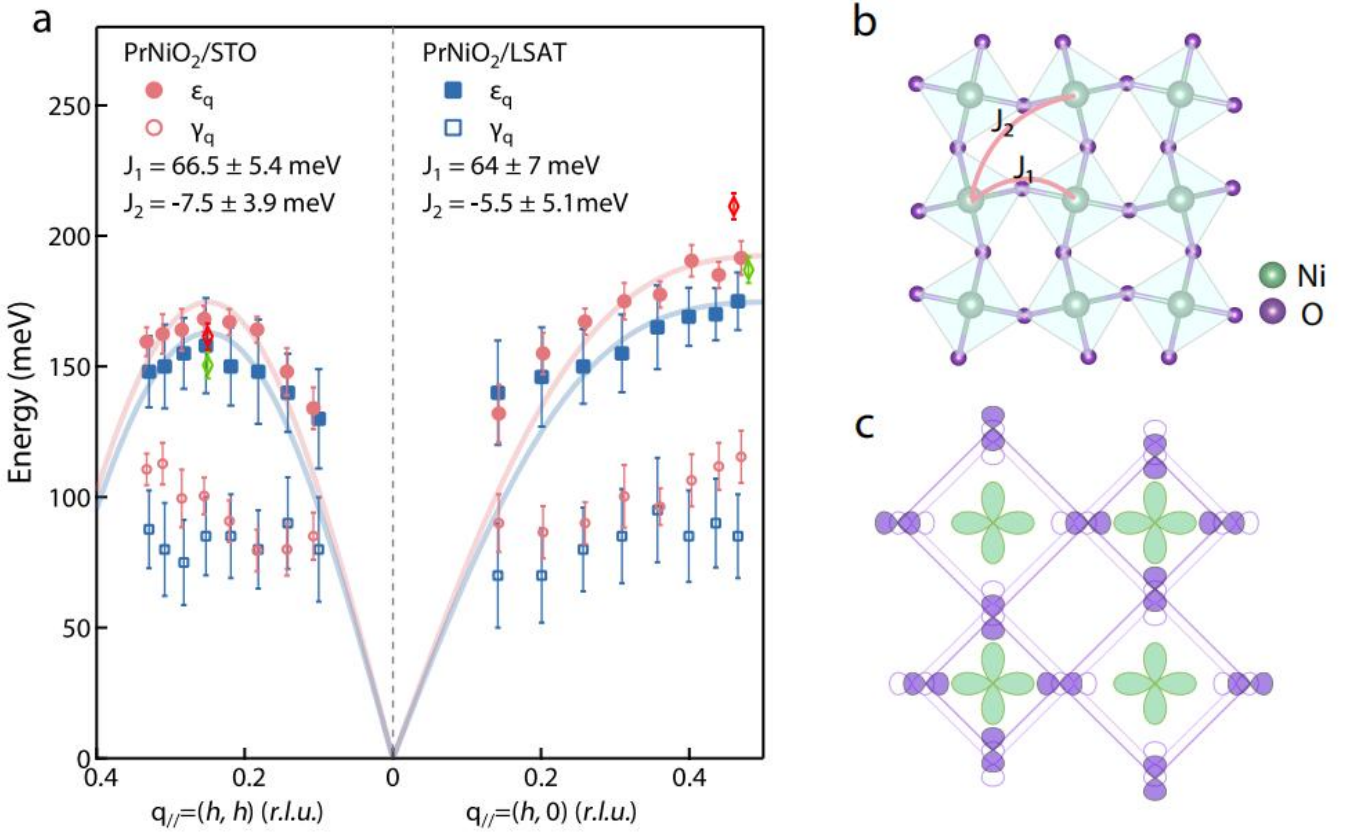

**FIG. 4: Dispersion of the magnetic excitations in PrNiO$_2$ and comparison to the model calculations.**

a, The mode energy ($\varepsilon_q$) and damping factor ($\gamma_q$) in PrNiO$_2$ grown on STO and LSAT. The red (green) diamond dots are $\varepsilon_q$ obtained from fitting the RIXS spectra of PrNiO$_2$ grown on STO (LSAT) on the zone boundaries measured by $\pi$ polarization. The solid lines represent best fit for the model of the spin 1/2 Heisenberg antiferromagnet on the square lattice using linear spin-wave theory. $J_1$ and $J_2$ represent the nearest-neighbour (NN) and next-nearest-neighbour (NNN), respectively. Error bars are estimated from the standard deviation obtained by the least-squares fitting algorithm and multiple times of fittings. b, Illustration of possible structural distortions in the PrNiO$_2$ films. c, Schematic plot of the bond-stretching phonon modes that may suppress the superexchange coupling strength.